\documentclass[showpacs,prl,twocolumn,aps]{revtex4}
\usepackage[T1]{fontenc}
\usepackage[latin1]{inputenc}
 \usepackage{bm}
 \usepackage{amsmath}
 \usepackage{amssymb}
 \usepackage{latexsym}
 \usepackage{amsfonts}
 \usepackage{epsfig}
 \usepackage{psfrag}

 \newcommand{\ket}[1]{\left|#1\right>}
 \newcommand{\bra}[1]{\left<#1\right|}

 \newcommand{\f}[1]{\mbox{\boldmath$#1$}}

 \newcommand{\bea}{\begin{eqnarray}}
 \newcommand{\ea}{\end{eqnarray}}
 \newcommand{\eea}{\end{eqnarray}}
 \newcommand{\ord}{{\cal O}}

\begin{document}

\title{Sweeping from the superfluid to Mott phase in the Bose-Hubbard model}

\author{Ralf Sch\"utzhold$^{1,*}$, Michael Uhlmann$^{1}$, Yan Xu$^{1}$,
and Uwe R.~Fischer$^{2,\dagger}$}

\affiliation{$^{1}$Institut f\"ur Theoretische Physik,
Technische Universit\"at Dresden, D-01062 Dresden, Germany
\\
$^{2}$Eberhard-Karls-Universit\"at T\"ubingen,
Institut f\"ur Theoretische Physik\\
Auf der Morgenstelle 14, D-72076 T\"ubingen, Germany}

\begin{abstract}
We study the sweep through the quantum phase transition from the
superfluid to the Mott state for the Bose-Hubbard model with a
time-dependent tunneling rate $J(t)$.
In the experimentally relevant case of exponential decay,
$J(t)\propto e^{-\gamma t}$, an adapted mean-field expansion for large
fillings~$n$ yields a scaling solution for the fluctuations.
This enables us to analytically calculate the evolution of the number and
phase variations (on-site) and correlations (off-site) for slow
($\gamma\ll\mu$), intermediate, and fast (non-adiabatic $\gamma\gg\mu$)
sweeps, where $\mu$ is the chemical potential.
Finally, we derive the dynamical decay of the off-diagonal long-range
order as well as the temporal shrinkage
of the superfluid fraction  in a persistent ring-current setup.
\end{abstract}

 
\maketitle

For many systems, the equilibrium properties (the thermal ensemble at
finite temperatures or the ground state at zero temperature) are quite
well understood.
However, strictly speaking, the equilibrium state applies to purely
static situations only.
For dynamical systems, the adiabatic theorem states that the actual
quantum state remains close to the ground state (at zero temperature)
if the external time-dependence is slow enough, i.e., much slower than
the internal frequencies of the system determined by its energy gaps.
If this adiabaticity condition fails, however, non-equilibrium effects
become important, leading to many intriguing phenomena.
For example, the usual split into the ground state plus small quantum
fluctuations around it is no longer unique, which usually leads to
effects such as the amplification of quantum fluctuations and the
creation of quasi-particles.

\begin{figure}[hbt]
\centerline{\mbox{\epsfxsize=5.25cm\epsffile{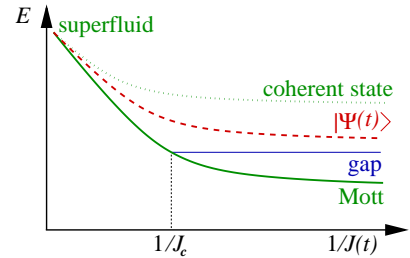}}}
\caption{[Color online] Sketch (not to scale) of level structure near
 the critical
 point, i.e., plot of the energy of the ground state (green solid
 line) and the energy gap (thin blue line) as a function of the
 external parameter~$1/J(t)$.
 Since the response time diverges at the critical point $J=J_c$,
 adiabaticity breaks down near the transition for any finite sweep
 velocity~$dJ/dt$.
 Hence the actual state $|\Psi(t)\rangle$ (red dashed line) slightly
 deviates from the exact ground state already before the critical
 point $J=J_c$ is reached and finally lies energetically somewhere
 between the coherent state (dotted green line) and the Mott state.}
\label{pic}
\end{figure}

A prototypical example for such a situation is a second-order quantum
phase transition \cite{SachdevBook}, cf.~Fig.\,\ref{pic}.
At the critical point, the energy gap vanishes and the response time
diverges.
Consequently, sweeping through the phase transition by means of a
time-dependent external parameter $1/J(t)$ with a finite velocity
$dJ/dt<0$ inevitably violates adiabaticity close enough to the
critical point and so generates non-equilibrium effects.
Fostered by the tremendous progress in the experimental capabilities,
there has been increasing interest in quantum phenomena at low
temperatures in general and quantum criticality in particular.
In view of the creation of entanglement \cite{VidalKitaev,Dorner},
quantum phase transitions are also relevant for quantum information
(the continuum limit of an adiabatic quantum algorithm
represents just a sweep through a quantum phase transition).
Finally, dynamical quantum phase transitions bear strong similarities
to cosmological phenomena:
If the stable phase permits topological defects
(symmetry-breaking transition), a quantum version of the Kibble-Zurek
mechanism may create those defects
\cite{Damski,ZDZ,Dziarmaga,DamskiZurek,Cucchietti}.
In case the initial spectrum contains gapless modes,
their quantum fluctuations will generally be amplified in close
analogy to cosmic inflation \cite{Ralf,inflation}.
Both phenomena can be understood in terms of the emergence of an
effective cosmic horizon which corresponds to the loss of causal
connection and the breakdown of adiabaticity near the critical point.

Besides the Ising model
(which can be solved analytically \cite{SachdevBook,Dziarmaga}),
another archetypical example is the quantum phase transition between
the superfluid and the Mott insulator state in the Bose-Hubbard model
(where no closed solution has been found).
This model provides a simplified description for the problem of
interacting bosons hopping on a lattice \cite{BoseHubbard};
its implementation with atoms/molecules in optical lattices
\cite{BosonicMott}, and the subsequent experimental realization
of the Mott-Hubbard transition \cite{MottBEC} has recently caused a
large amount of research activity \cite{HubbardToolBox}.
The superfluid phase possesses a rich spectrum ranging from
topological defects (e.g., kinks in one dimension) to gapless
quasi-particle excitations (phonons) but the Mott state does not.
Hence the (quantum) Kibble-Zurek mechanism requires the dynamical
transition from the Mott to the superfluid state (quantum melting)
whereas the amplification of the quantum fluctuations of the phonon
modes occurs in the opposite direction (quantum freezing).
The (mostly numerical) studies thus far
\cite{Cucchietti,Polkovnikov}
have been devoted to the first case (quantum melting).
Here, we study the dynamical behavior of the superfluid to Mott
quantum phase transition starting, by contrast to previous
investigations, from the superfluid side of the transition.
We thus investigate the quantum freezing of number fluctuations
\cite{note,Gerbier} and predict their dependence on the sweep
rate~$\gamma$.
The thereby created number-squeezed state and its
number fluctuations can be studied experimentally,
for example by the interference of condensates released from different
optical lattice wells \cite{Orzel}.
As further observables representing the actual quantum state
$\ket{\Psi(t)}$, to bring out the difference to either
the superfluid or Mott states (cf.~Fig.\,\ref{pic}),
we derive the temporal evolution of the off-diagonal long-range order
and the superfluid fraction.

The Bose-Hubbard model with local contact interactions is, on a rather
general type of lattice, described by the Hamiltonian
\bea
\label{Hamiltonian}
\hat H
=
J(t)\sum\limits_{\alpha\beta}
M_{\alpha\beta}\hat a_\alpha^\dagger\hat a_\beta
+\frac{U}{2}\sum\limits_\alpha(\hat a_\alpha^\dagger)^2\hat a_\alpha^2
\,. \label{Hubbard}
\ea
Here $\hat a_\alpha^\dagger$ and $\hat a_\alpha$ denote the bosonic
creation and annihilation operators at the lattice site $\alpha$.
The energy scale is set by the on-site repulsion $U>0$ and the
time-dependent tunneling rate $J(t)\geq0$.
The structure of the lattice is encoded in the matrix
$M_{\alpha\beta}$, which is supposed to obey the same
(translational) symmetry group, e.g., for a one-dimensional chain with
nearest-neighbor hopping, we have
$M_{\alpha\beta}=\delta_{\alpha,\beta}
-\frac12(\delta_{\alpha,\beta+1}+\delta_{\alpha,\beta-1})$.

The initial state is the homogeneous superfluid phase,
$J\gg J_c=\ord(U/n)$, see \cite{Penna},
with a large integer filling $n=\langle\hat n_\alpha\rangle\gg1$
(where $\hat n_\alpha=\hat a_\alpha^\dagger\hat a_\alpha$),
which facilitates a particle-number conserving mean-field expansion
\cite{Particle,castin}
\bea
\label{mean-field}
\hat a_\alpha
=
\left(\psi_0+\hat\chi_\alpha+\hat\zeta_\alpha\right)\hat A/\sqrt{\hat N}
\,,
\ea
with
$\hat A=\hat a_\Sigma(\hat a_\Sigma^\dagger\hat a_\Sigma)^{-1/2}\hat N^{1/2}$
and
$\hat A^\dagger\hat A=\hat N=\sum_\alpha\hat n_\alpha$,
where
$\hat a_\Sigma=\sum_\alpha\hat a_\alpha$.
Here the main idea is to expand the original operator $\hat a_\alpha$
into powers of $n\gg1$, considering the (formal) limit
$n\uparrow\infty$ with the chemical potential $\mu=Un$ remaining
finite, i.e., $U=\ord(1/n)$.
The leading term is the order parameter $\psi_0=\ord(\sqrt{n})$ and
the quantum corrections $\hat\chi_\alpha=\ord(n^0)$ correspond to
quasi-particle excitations.
For the validity of this expansion, the remaining higher-order
corrections $\hat\zeta_\alpha$ must be small
$\hat\zeta_\alpha\ll\ord(n^0)$.
Inserting the above expansion~(\ref{mean-field}) into the equation
of motion $i\hbar\partial_t\hat a_\alpha=
J(t)\sum_{\beta}M_{\alpha\beta}\hat a_\beta+
U\hat n_\alpha\hat a_\alpha$ obtained from (\ref{Hamiltonian}) and
sorting into powers of $n$ yields the analogue of the
Bogoliubov-de~Gennes equations for the excitations at site $\alpha$
($\hbar \equiv 1$ throughout)
\bea
\label{Bogoliubov-deGennes}
i\partial_t\hat\chi_\alpha
=
J(t)\sum\limits_{\beta}M_{\alpha\beta}\hat\chi_\beta
+2U|\psi_0^2|\hat\chi_\alpha+U\psi_0^2\hat\chi_\alpha^\dagger
\,,
\ea
and for the remaining higher-order corrections
\bea
\label{higher-order}
i\partial_t\hat\zeta_\alpha
&=&
J(t)\sum\limits_{\beta}M_{\alpha\beta}\hat\zeta_\beta
+2U|\psi_0^2|\hat\zeta_\alpha+U\psi_0^2\hat\zeta_\alpha^\dagger+
\\
&&
+2U\psi_0\hat\chi_\alpha^\dagger\hat\chi_\alpha
+U\psi_0^*\hat\chi_\alpha^2+U\hat\chi_\alpha^\dagger\hat\chi_\alpha^2
+\ord(U\hat\zeta_\alpha)
\nonumber
\,.
\ea
Deep in the superfluid phase (our initial state), the higher-order
corrections $\hat\zeta_\alpha$ are small and the mean-field
expansion~(\ref{mean-field}) works very well.
If we approach the Mott phase, however, these corrections start to
grow according to Eq.\,(\ref{higher-order}) and, at some point, the
mean-field expansion~(\ref{mean-field}) breaks down.
The characteristic time-scale of this breakdown can be estimated from
the nonlinear source terms in Eq.\,(\ref{higher-order}) which are
suppressed to $\ord(1/\sqrt{n})$ in view of $U=\ord(1/n)$,
$\psi_0=\ord(\sqrt{n})$ and $\hat\chi_\alpha=\ord(n^0)$.
Hence (starting in the superfluid phase), the higher-order corrections
remain small as long as ${Ut\sqrt{n}\ll1}$, i.e., for evolution times
of order $t=\ord(\sqrt{n})$.
Thus we can extra\-polate the mean-field expansion even into the Mott
phase for some time $t=\ord(\sqrt{n})$, and follow the evolution of
the instabilities which develop because the superfluid state is no
longer the ground state of the system.

The polar decomposition
$\hat a_\alpha=\exp\{i\hat\phi_\alpha\}\sqrt{\hat n_\alpha}$
yields the linearized number fluctuations $\delta\hat n_\alpha$
via $\hat n_\alpha=n+\delta\hat n_\alpha$ and accordingly the
conjugate phase fluctuations $\delta\hat\phi_\alpha$, where
$\hat\chi_\alpha=
\psi_0[\delta\hat n_\alpha/(2n)+i\delta\hat\phi_\alpha]
+\ord(1/\sqrt{n})$.
In terms of these fluctuations $\delta\hat n_\alpha$,
Eq.\,(\ref{Bogoliubov-deGennes}) can be diagonalized by a normal-mode
expansion into the eigenvectors of the matrix $M_{\alpha\beta}$
labeled by the generalized momenta $\kappa$
\bea
\left(
\frac{\partial}{\partial t}\,
\frac{1}{J(t)}\,\frac{\partial}{\partial t}
+8\lambda_\kappa
\left[Un+2J(t)\lambda_\kappa\right]
\right)
\delta\hat n_\kappa
=0
\,,\label{normalmodeEq}
\ea
where $\lambda_\kappa$ are the corresponding eigenvalues of
$M_{\alpha\beta}$.

For small $\lambda_\kappa$, the above evolution equation is analogous
to the modes of a quantum field within an expanding
universe with the wavenumber $k\propto\sqrt{\lambda_\kappa}$.
Pursuing the similarity a bit further, we get the analogue of a cosmic
horizon with the horizon size
\bea
\Delta_{\rm h}(t)=\int\limits_t^\infty dt'\,\sqrt{J(t')Un}
\,,\label{horizon}
\ea
if $J(t)$ changes fast enough \cite{Ralf,inflation}.
Note that the upper integral limit should not be taken literally,
since it must
still be within the region of validity of the mean-field expansion.
In case an effective horizon occurs, its size constantly decreases.
Hence all modes (with small $\lambda_\kappa$) will qualitatively
follow the same evolution -- but at different times:
Initially, the wavelength of the modes is well inside the horizon,
$\lambda_\kappa\Delta_{\rm h}^2\gg1$, and the modes oscillate almost
freely.
At some point of time, however, the constantly shrinking horizon
closes in, $\lambda_\kappa\Delta_{\rm h}^2=\ord(1)$, and thus the causal
connection across a wavelength is lost.
After that, the modes cannot oscillate anymore and freeze.
This process leads to the amplification of the initial quantum
fluctuations via squeezing in analogy to the inflationary
epoch in the early universe \cite{inflation,Ralf}.
We note that these considerations show a major weakness of the
adiabatic--impulse approximation  used in \cite{DamskiZurek}, since
modes with different $\lambda_\kappa$ become non-adiabatic at
different times.

Since the tunneling rate $J$ depends exponentially on the amplitude of
the laser which generates the optical lattice \cite{essentially},
we shall consider an exponential time-dependence
$J(t)=J_0\exp\{-\gamma t\}$ in the following, which is implying the
emergence of a quasiparticle horizon according to Eq.\,(\ref{horizon}).
Furthermore, Eq.\,(\ref{normalmodeEq}) reveals that $\lambda_\kappa$
can be absorbed by a suitable redefinition of the time coordinate in
this case, $\tau_\kappa=-4\lambda_\kappa J(t)/\gamma$, leading to
\bea
\left(
\frac{\partial^2}{\partial\tau_\kappa^2}
+\left[1-\frac{2}{\tau_\kappa}\,\frac{Un}{\gamma}\right]
\right)
\delta\hat n_\kappa
=0
\,.\label{scalingEq}
\ea
The universal scaling solution resulting from the above equation
confirms that modes with different $\lambda_\kappa$ display the same
behavior, but at different times.
The only remaining dimensionless parameter is the ratio
$\nu=Un/\gamma$ which is a measure of the sweep velocity:
$\nu\gg1$ implies a slow and $\nu\ll1$ a fast (non-adiabatic) sweep.
The differential equation (\ref{scalingEq}) has a universal scaling
solution in terms of Whittaker functions \cite{Abramowitz}
\bea
\delta\hat n_\kappa
=
\sqrt{n}\,e^{-\pi\nu/2}\,W_{i\nu,1/2}(2i\tau_\kappa)\,\hat b_\kappa
+{\rm h.c.}
\ea
The quasi-particle operator $\hat b_\kappa$ annihilates the adiabatic
ground state $\hat b_\kappa\ket{\rm in}=0$ at early times
$\tau_\kappa\downarrow-\infty$, where the modes oscillate like
$e^{\pm i\tau_\kappa}$ (which can also be obtained from the
asymptotic behavior of the Whittaker functions).
According to the arguments presented after Eq.\,(\ref{higher-order}),
the mean-field expansion remains valid for intermediate times
with ${Ut\sqrt{n}\ll1}$ but $\gamma t\gg1$, and hence can be
extrapolated to the ``late-time'' regime $\tau_\kappa\uparrow0$ where
the functions $W_{i\nu,1/2}(2i\tau_\kappa)$ approach a constant value,
i.e., the number fluctuations freeze ($J\downarrow0$).
Due to the perfect scaling solution, the frozen value is independent
of $\kappa$, but the decaying corrections do depend on $\lambda_\kappa$:
\bea
\langle\delta\hat n_\kappa^2\rangle
\equiv
\bra{{\rm in}} (\delta\hat n_\kappa)^2 \ket{{\rm in}}
=
n\frac{1-e^{-2\pi\nu}}{2\pi\nu}
+\ord(t\lambda_\kappa e^{-\gamma t})
\,.
\ea
Since the leading term is independent of $\kappa$, it just yields a
local ($\propto\delta_{\alpha,\beta}$) contribution after the mode sum
($\kappa\to\alpha$) and thus leads to frozen on-site number variations
\bea
\Delta^2(n_\alpha)
=
\langle\hat n_\alpha^2\rangle-
\langle\hat n_\alpha\rangle^2
=
\langle\delta\hat n_\alpha^2\rangle
=
n\frac{1-e^{-2\pi\nu}}{2\pi\nu}
\,.
\ea
For a rapid sweep $\nu\ll1$, this variation $\Delta^2(n_\alpha)$
approaches a constant value $n$, which is characteristic of a
superfluid phase with Poissonian number statistics \cite{Javanainen}.
For a slow  sweep $\nu\gg1$, the variation $\Delta^2(n_\alpha)$
becomes small, $\propto n/\nu$, and so approaches the behavior deep
into the Mott phase (with $J\downarrow 0$),
where $\Delta^2(n_\alpha)=0$, cf.~\cite{Schroll}.
For $J\downarrow 0$, the energy $\langle\hat H\rangle$ is basically
determined by the variation $\Delta^2(n_\alpha)$ and hence this
quantity describes the energetic location of the final state in
comparison to the coherent state and the Mott state, see
Fig.\,\ref{pic}. 

The sub-leading corrections in $\langle\delta\hat n_\kappa^2\rangle$
depend on $\kappa$ and hence determine the off-site ($\alpha\neq\beta$)
number correlations which decay exponentially for $\gamma t\gg1$
\bea
\langle\hat n_\alpha\hat n_\beta\rangle-
\langle\hat n_\alpha\rangle\langle\hat n_\beta\rangle
=
\langle\delta\hat n_\alpha\delta\hat n_\beta\rangle
=
\ord(\gamma t\,e^{-\gamma t})
\,,
\ea
where we have omitted finite-size effects (which scale with the
inverse number of lattice sites), since strictly speaking the mode sum
does not include the zero-mode $\kappa=0$.

The conjugate phase fluctuations can be derived in an analogous manner
and are determined by the derivatives of the Whittaker functions
$dW_{i\nu,1/2}/d\tau_\kappa$.
As one would expect \cite{note}, they do not freeze -- but increase:
\bea
\langle\delta\hat\phi_\kappa^2\rangle
=
\nu\frac{1-e^{-2\pi\nu}}{2\pi n}\,\gamma^2t^2
+\ord(\gamma t\ln\lambda_\kappa)
\,. \label{phasefluct} 
\ea
Again the leading (first) term is independent of $\kappa$ and thus
yields the on-site fluctuations
$\langle\delta\hat\phi_\alpha^2\rangle$ only
(which generate the quadratically growing quantum depletion
$\langle\hat\chi_\alpha^\dagger\hat\chi_\alpha\rangle$
and anomalous term $\langle\hat\chi_\alpha^2\rangle$).
The second term is the leading $\kappa$-dependent contribution and
determines the off-site phase correlations
$\langle\delta\hat\phi_\alpha\delta\hat\phi_\beta\rangle$.

The ascent of the phase fluctuations can be understood as a
consequence of the emergence of an effective horizon which
entails the loss of causal connection between different sites and thus
the decay of the phase coherence across the lattice.
As one interesting observable, let us discuss the evolution of the
off-diagonal long-range order between sites $\alpha$ and $\beta$
defined by the correlator
$\langle\hat a_\alpha^\dagger(t)\hat a_\beta(t)\rangle$.
Using the mean-field results valid at intermediate times with
$\gamma t\gg1$ but $Ut\sqrt{n}\ll1$ as the initial conditions, we may
derive the ensuing stages of the quantum evolution, where the
tunneling rate $J(t\gg1/\gamma)\lll1$ is exponentially small and can
be neglected.
In this limit, the evolution of the operators can be approximated by
$d\hat a_\alpha/dt=-iU\hat n_\alpha\hat a_\alpha$
which possesses the simple solution
$\hat a_\alpha(t)=\exp\{-iU\hat n_\alpha^0 t\}\hat a_\alpha^0$.
Consequently, we obtain
$\langle\hat a_\alpha^\dagger(t)\hat a_\beta(t)\rangle
=n\langle\exp\{iU(\hat n_\alpha-\hat n_\beta)t\}\rangle
+\ord(\sqrt{n})$.
The frozen first-order number fluctuations $\delta\hat n_\alpha$
are in a squeezed state which can (for $n\gg1$)
be approximated by a (continuous) Gaussian distribution.
For a Gaussian variable $X$ with $\langle X\rangle=0$,
the exponential average yields
$\langle\exp\{iX\}\rangle=\exp\{-\langle X^2\rangle/2\}$
and hence we get
\bea
\label{order}
\langle\hat a_\alpha^\dagger(t)\hat a_\beta(t)\rangle
\approx
n
\exp\{-U^2t^2\Delta^2(n_\alpha)\}
\,.
\ea
Note that this expression is only valid for time scales
$1/\gamma \ll t \ll 1/U$; for $Ut\in2\pi\mathbb N$ all the exponential
factors equal unity again and we have a revival of the phase
coherence and thus long-range order similar to a spin echo.
(Note that we omit the coupling to the environment in our derivation.)
%
Furthermore, the above Gaussian decay is independent of the distance
between the sites $\alpha$ and $\beta$.
Since the Fourier transform of
$\langle\hat a_\alpha^\dagger(t)\hat a_\beta(t)\rangle$
determines the structure factor $S(\f{k})$, the decay of the
long-range order (\ref{order}) directly corresponds to the temporal
decrease of the peak in $S(\f{k})$ at $\f{k}=0$.
This dependence on time and sweep rate can be observed with
time-of-flight measurements \cite{GerbierII}
(which basically map out $|S(\f{k})|$) by varying $\gamma$ and the
time delay between the phase transition and the release of the
condensate.

Let us calculate another simple experimental signature, which involves
just neighboring sites, namely the superfluid fraction $n_{\rm sf}/n$.
To this end, we specify our lattice and choose a one-dimensional
ring-geometry $\alpha\to\ell$ with the circumference coordinate $\ell$
\cite{ring-current}.
In order to generate a persistent current, we impose a small phase
gradient externally by trapping a flux quantum on the ring:
$\psi_0=\sqrt{n}\exp\{-i\mu t+2\pi i\ell/L\}$, where ${L\gg1}$ denotes
the number of sites.
Naturally, for a decreasing tunneling rate $J(t)$, the induced current
diminishes also.
In addition, the number of particles contributing to this flux goes
down -- which will be used as a measure for the superfluid fraction.
The flux can be determined by the usual Noether current corresponding
to the $U(1)$ invariance and is related to the nearest
neighbor correlation function
$\langle\hat a_\ell^\dagger\hat a_{\ell+1}\rangle$.
Hence, the shrinkage of the superfluid fraction is given by the
same \cite{footnote} expression as in Eq.\,(\ref{order})
\bea
\label{fraction}
\frac{n_{\rm sf}}n
\approx
\exp\left\{-U^2t^2
n\frac{1-e^{-2\pi\nu}}{2\pi\nu}
\right\}
\,.
\ea
In the case of a very rapid (impulse) sweep, $\nu\ll1$, the superfluid
fraction decays with $\exp\{-nU^2t^2\}$, i.e., independent of
$\gamma$.
Conversely, for a slow sweep $\nu\gg1$, the decay takes much
longer: $\exp\{-nU^2t^2/(2\pi\nu)\}$.

In conclusion, an adapted mean-field expansion enables us to analytically
calculate the freezing of number fluctuations and the growth of phase
fluctuations during an (exponential) sweep through the
superfluid $\to$ Mott quantum phase transition.
As further experimental signatures, we predict the temporal decay of
the superfluid fraction $n_{\rm sf}/n$ and of the central $\f{k}=0$ peak
in the structure factor $S(\f{k})$.

This work was supported by the Emmy Noether Programme of the
German Research Foundation (DFG) under grant No.~SCHU~1557/1-1,2.
The authors also gratefully acknowledge the COSLAB
Programme of the ESF.


$^*$\,{\footnotesize\sf schuetz@theory.phy.tu-dresden.de}
$^\dagger$\,{\footnotesize \sf uwe.fischer@uni-tuebingen.de}


\begin{thebibliography}{9999}
\bibitem{note}
In the superfluid phase, the phase fluctuations are small and the
number fluctuations are large [for $J \gg U$ in Eq.\,(\ref{Hubbard})
we approximately have a coherent state with Poissonian number
statistics],   
whereas the Mott phase shows the opposite behavior.
\bibitem{footnote}
Note that the coincidence between the long and short-range behavior in
Eqs.\,(\ref{order}) and (\ref{fraction}) is a consequence of the
exponential time-dependence of $J(t)$ which allows for a scaling
solution.
%
The freezing of the number fluctuations and the growth of the phase
fluctuations is a general phenomenon, but the dependence on $\kappa$
is different for other sweep dynamics; e.g., for a linear sweep
$dJ/dt= $ const., the leading contributions to (\ref{phasefluct}) 
would scale as
$\lambda_\kappa^{-1/3}$ for small $\lambda_\kappa$ \cite{Ralf}.
\end{thebibliography}
\end{document}